\documentclass[aps,prl,twocolumn,superscriptaddress,showpacs,10pt]{revtex4-2}
\usepackage{graphicx}% Include figure files
\usepackage{dcolumn}% Align table columns on decimal point
\usepackage{bm}% bold math
\usepackage{upgreek}
\usepackage{color}
\usepackage{soul}
\usepackage{epstopdf}
\usepackage{epsfig}
\usepackage{units}
\usepackage{longtable}
\usepackage{floatrow}
\usepackage{verbatim} %In the preamble 
\usepackage{latexsym}
\usepackage{textcomp,gensymb}
\usepackage{amssymb}
\usepackage{amsmath}
\usepackage[normalem]{ulem}
\usepackage{lineno}
\usepackage{soul}
\usepackage{hyperref}% add hypertext capabilities
\usepackage{xcolor}
\DeclareUnicodeCharacter{308}{}

\makeatletter
\newcommand{\fmarki}{*}
\newcommand{\fmarkii}{*}
\newcommand{\fmarkiii}{*}
\newcommand{\fmarkiv}{\ensuremath{\mathsection}}
\newcommand{\fmarkv}{\ensuremath{\mathparagraph}}
\newcommand{\fmarkvi}{\ensuremath{\|}}
\newcommand{\fmarkvii}{**}
\newcommand{\fmarkviii}{\ensuremath{\dagger\dagger}}
\newcommand{\fmarkix}{\ensuremath{\ddagger\ddagger}}
                
\def\@fnsymbol#1{{\ifcase#1\or \fmarki\or \fmarkii\or \fmarkiii\or \fmarkiv\or \fmarkv\or \fmarkvi\or \fmarkvii\or \fmarkviii\or \fmarkix \else\@ctrerr\fi}}
\makeatother

\newcommand{\Fig}[1]{{Fig.\ \ref{#1}}}

\begin{document}

\title{Interfering trajectories in a ballistic Andreev cavity}

\author{Pankaj Mandal}
\email{pankaj.mandal@physik.uni-wuerzburg.de}
\affiliation{Faculty for Physics and Astronomy (EP3),
Universit\"at W\"urzburg, Am Hubland, D-97074, W\"urzburg, Germany}
\affiliation{Institute for Topological Insulators, Am Hubland, D-97074, W\"urzburg, Germany}

\author{Marcel Kaschper}
\affiliation{Faculty for Physics and Astronomy (EP3),
Universit\"at W\"urzburg, Am Hubland, D-97074, W\"urzburg, Germany}
\affiliation{Institute for Topological Insulators, Am Hubland, D-97074, W\"urzburg, Germany}

\author{Fernando Dominguez}
\affiliation{W\"urzburg-Dresden Cluster of Excellence ct.qmat, Universit\"at W\"urzburg, D-97074 W\"urzburg, Germany}
\affiliation{Institut f\"ur Theoretische Physik und Astrophysik, Universit\"at W\"urzburg, D-97074 W\"urzburg, Germany}

\author{Soumi Mondal}
\affiliation{Faculty for Physics and Astronomy (EP3),
Universit\"at W\"urzburg, Am Hubland, D-97074, W\"urzburg, Germany}
\affiliation{Institute for Topological Insulators, Am Hubland, D-97074, W\"urzburg, Germany}

\author{Lukas Lunczer}
\affiliation{Faculty for Physics and Astronomy (EP3),
Universit\"at W\"urzburg, Am Hubland, D-97074, W\"urzburg, Germany}
\affiliation{Institute for Topological Insulators, Am Hubland, D-97074, W\"urzburg, Germany}

\author{Dongyun Chen}
\affiliation{Faculty for Physics and Astronomy (EP3),
Universit\"at W\"urzburg, Am Hubland, D-97074, W\"urzburg, Germany}
\affiliation{Institute for Topological Insulators, Am Hubland, D-97074, W\"urzburg, Germany}

\author{Martin P. Stehno}
\affiliation{Faculty for Physics and Astronomy (EP3),
Universit\"at W\"urzburg, Am Hubland, D-97074, W\"urzburg, Germany}
\affiliation{Institute for Topological Insulators, Am Hubland, D-97074, W\"urzburg, Germany}

\author{Ewelina M. Hankiewicz}
\affiliation{W\"urzburg-Dresden Cluster of Excellence ct.qmat,
Universit\"at W\"urzburg, D-97074 W\"urzburg, Germany}
\affiliation{Institut f\"ur Theoretische Physik und Astrophysik,
Universit\"at W\"urzburg, D-97074 W\"urzburg, Germany}

\author{Bj\"orn Trauzettel}
\affiliation{W\"urzburg-Dresden Cluster of Excellence ct.qmat,
Universit\"at W\"urzburg, D-97074 W\"urzburg, Germany}
\affiliation{Institut f\"ur Theoretische Physik und Astrophysik,
Universit\"at W\"urzburg, D-97074 W\"urzburg, Germany}

\author{Teun M. Klapwijk}
\affiliation{Faculty for Physics and Astronomy (EP3),
Universit\"at W\"urzburg, Am Hubland, D-97074, W\"urzburg, Germany}
\affiliation{Retired at Kavli Institute of NanoScience, Faculty of Applied Sciences, Delft University of Technology, Lorentzweg 1, 2628 CJ Delft, The Netherlands}

\author{Charles Gould}
\email{gould@physik.uni-wuerzburg.de}
\affiliation{Faculty for Physics and Astronomy (EP3),
Universit\"at W\"urzburg, Am Hubland, D-97074, W\"urzburg, Germany}
\affiliation{Institute for Topological Insulators, Am Hubland, D-97074, W\"urzburg, Germany}

\author{Laurens W. Molenkamp}
\email{molenkamp@physik.uni-wuerzburg.de}
\affiliation{Faculty for Physics and Astronomy (EP3),
Universit\"at W\"urzburg, Am Hubland, D-97074, W\"urzburg, Germany}
\affiliation{Institute for Topological Insulators, Am Hubland, D-97074, W\"urzburg, Germany}

\date{\today}

\begin{abstract}
The conventional description of transport through the interface between a normal conductor and a superconductor reduces the system to a one-dimensional problem treating Andreev reflection based on a zero-dimensional Sharvin type point-contact model, and effectively neglects all considerations of device geometry. While this has been successful in systems where conductance in the normal material is in the diffusive transport regime, such an over-simplification of the problem fails in other transport regimes. In particular, when transport is ballistic as in a typical semiconductor-superconductor hybrid structure, geometrical effects are inherently important, and a proper description must consider a one-dimension contact injecting into a two-dimensional ballistic cavity. We present the first study of this regime and explore the bias-voltage dependence of Andreev transport in a cavity-type device comprised of a high mobility HgTe quantum well side-contacted by one superconducting and one normal contact, each creating a one-dimensional interface. The enhanced conductance from Andreev transport features two finite bias conductance peaks,  observed at energies within the energy gap of the superconductor. 
 Interestingly, these two peaks respond differently to the application of a perpendicular-to-plane magnetic field. Using a semi-classical model for the quantum transport within the cavity, we are able to attribute each peak to a different class of ballistic trajectories. One class is dominated by normal reflection, and its interference condition is independent of magnetic field, whereas the other one contains retro-reflected Andreev processes at the superconductor interface. These create closed trajectories that are strongly suppressed by magnetic field due to Aharonov-Bohm and Doppler shift effects.

\end{abstract}

\maketitle

\section*{Introduction}

Andreev reflection\cite{Andreev1964} is a fundamentally important mechanism of charge transport at a normal-metal\textemdash superconductor interface. The first observation of excess current\cite{taylor1963excess} in 1963, triggered extensive experimental work\cite{notarys1973proximity, akimenko1976ac, klapwijk1977regimes, octavio1978nonequilibrium, gubankov1981excess} on normal-metal\textemdash superconductor structures. These explorations culminated in the seminal paper by Blonder, Tinkham and Klapwijk (BTK) \cite{blonder1982transition} that established the picture through which these devices continue to be viewed even today. Their model described a Sharvin type zero-dimensional point contact-like injection of an electron from the normal metal into the superconductor. It thus reduces the device to a single point contact, ignoring the normal metallic contact.    

In recent years, it has been suggested that combining superconductors with material of non-trivial band topology could exhibit exotic phenomena such as a Majorana state \cite{fu2008superconducting, qi2010chiral} or  triplet superconductivity\cite{ van2011spin}. As many of these materials emerge from band inversion, they share properties with semiconductors, and in particular often have long electron mean free path, of the order of micrometers. This allows devices in the mesoscopic or ballistic regime. A proper description of such devices requires considering ballistic transport from one lead to the other through a cavity, and thus necessarily must take two interfaces into account, which cannot be described as simple Sharvin type point contacts but must rather be treated as one-dimensional extended interfaces. 

Indeed, this concept is well appreciated in Josephson junctions, where the interference condition for trajectories bouncing between the two superconductors leads to multiple Andreev reflection and Andreev bound states \cite{rowell1968excess, octavio1983subharmonic}. The role of the second interface in the normal/superconductor case is however not yet explored. Pandey et al.,\cite{pandey2019andreev} consider a second interface in a device comprised of a graphene cavity with one normal and one superconducting lead, but restrict themselves to the one dimensional transport across a point contact like injection typical of BTK, while at the same time neglecting the effect of dynamical phase gain \cite{nazarov2009quantum} in the electron wavefunction as a result of its travel through the cavity. 

In this paper, we investigate a two-dimensional ballistic cavity made of a high mobility HgTe quantum well side contacted by both a normal-metal and a superconductor through extended one-dimensional interfaces to study Andreev transport beyond the regime of Sharvin like point contacts. Current-voltage spectroscopy of the device reveals quantum interference phenomena associated with the two-dimensional nature of the cavity exhibiting a nontrivial magnetic field dependence which can only be understood by considering the interplay of dynamical and Andreev phase, as well as the two-dimensional nature of the device. The one-dimensional nature of the interfaces is fundamentally important. Given the current interest in exploring novel materials in combination with superconductors, our findings lay the groundwork to go beyond the conventional Sharvin-type point contact description of Andreev reflection and evaluate experimental results accordingly.

%--------------------------------------------------------------Fig1
\begin{figure}[ht]%
\includegraphics[width=\columnwidth]{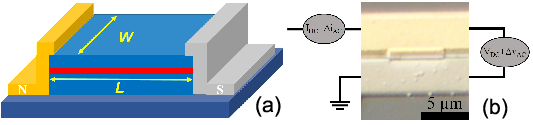}
\caption{(a) Schematic of the side-contacted device geometry. The HgTe quantum well (in red) is embedded between (Cd,Hg)Te layers (in blue) grown on a (Cd,Zn)Te substrate (in dark blue). $N$ and $S$ represent the normal contact (Au) and the superconducting contact (MoRe), respectively. The length ($L$) and width ($W$) of the device are indicated in the schematic. (b) Nomarski microscopy image of the device and simplified schematic of the measurement circuit.
}%
\label{Fig1}%
\end{figure}
%-------------------------------------------------------------------

\section*{Device and characterization} 

A schematic of the device is shown in \Fig{Fig1}(a). It features a rectangular mesa with a width (${W = 5.5\; \mu\text{m}}$) and length (${L = 1\; \mu\text{m}}$) patterned from a (Cd,Hg)Te/HgTe/(Cd,Hg)Te heterostructure, with the HgTe quantum well indicated in red. The mesa is pattered using wet-etching technique\cite{bendias2018high}, and its length is shorter than the electron mean free path ($l$) of $3\,\mu\text{m}$ (as determined from measurements on a Hall-bar made from the same material). It is contacted using a side-contact technology, \cite{bendias2018high, mandal2022finite} from one side with a normal gold contact, and from the other using superconducting MoRe. The measured critical temperature of the MoRe is ${\sim9.6\;\text{K}}$ which corresponds to a Bardeen–Cooper–Schrieffer (BCS) gap ($\Delta$) of 1.44 meV. 

A Nomarski microscopy image of the completed device is shown in \Fig{Fig1}(b), which also indicates the quasi four-probe measurement configuration used to exclude the effect of cryostat wiring in the electrical characterization. A bias voltage is applied by using a transformer to inductively add a small AC bias to the DC bias from a voltage source. This combined voltage is applied to a $214.8\;\text{k}\Omega$ series resistance to convert it to a current that is passed through a $1.9935\;\text{k}\Omega$ series reference resistor, and then into the junction using the Au contact, and onto ground at the MoRe contact. This current is the sum of a constant ($I_{DC}$) and a $13.37\;\text{Hz}$ oscillating ($\Delta I_{AC}$) component. The voltage falling over the junction is sampled using a voltage amplifier with a gain of 1000. The amplified signal is then fed to both a DC-voltmeter and a lock-in amplifier to detect the AC signal. This circuit allows a direct determination of the differential conductance (d$I$/d$V$) vs DC bias voltage $V_{b}$ across the sample.

The carrier density in the quantum well is ${\sim 7.8\times10^{11}\;  \text{cm}^{-2}}$, as extracted from the Shubnikov - de Haas period at large magnetic field (see Supplemental Material at \cite{supplementalmaterial} for section S1A). The corresponding Fermi level from established band structure calculations \cite{beugeling2025kdotpy} is 80 meV above the bottom of the conduction band, and thus much larger than the energy of the superconducting gap. The Fermi velocity ($v_{F}$) is ${ 7.7\times10^{5}\;  \text{m/s}}$, and the Fermi wavelength ($\lambda_{F}$) is $28\;\text{nm}$, which is much smaller than $L$.  In this highly n-type transport regime, the conductance is dominated by bulk carriers and the contribution of helical edge modes at the edges of the topological insulator can be neglected. 

\section*{Results} 
Figure 2 shows the differential conductance d$I$/d$V$, as a function of $V_{b}$ measured at base temperature ($\sim 25\;\text{mK}$) as well as at $13\;\text{K}$, above the critical temperature of MoRe. The 13 K measurement shows a linear relation between differential conductance and bias voltage, which can be attributed to a self-gating effect \cite{kristensen2000bias} that results in a small linear change in Fermi level as a function of bias voltage. For ease of presentation, in the rest of this manuscript, this  antisymmetric contribution to the differential conductance curves has been subtracted from the data, as shown in the inset of Fig. 2 as an example.

At 13 K, other than the self-gating, the differential conductance is featureless. Once cooled to well below the superconducting transition temperature of MoRe, excess conductance is observed at voltages $V_{b}$ below the superconducting gap of MoRe ($\Delta\approx1.4\; \text{meV}$). This implies a significant contribution of Andreev reflection \cite{Andreev1964} at the superconducting interface in this regime.

%--------------------------------------------------------------Fig2
\begin{figure}[ht]%\label{fig:conductance}%
\includegraphics[width=\columnwidth]{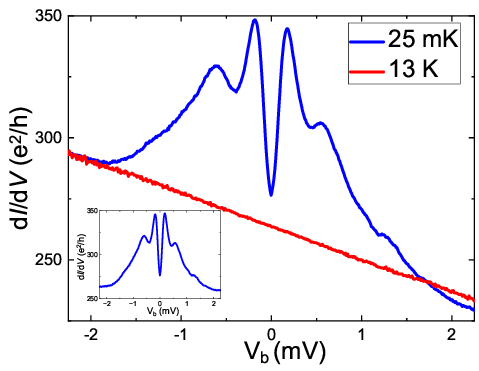}
\caption{ Bias-voltage dependence of d$I$/d$V$ for zero gate-voltage measured at 25 mK and 13 K. (Inset: Plot of base temperature d$I$/d$V$ after removing the antisymmetric background.)  }
\label{Fig2}%
\end{figure}
%------------------------------------------------------------------- 

The structure of this excess conductance is quite rich in features, with a sharp dip in conductance at zero bias-voltage and subgap peaks appearing at roughly 180 and $600\;\mu\text{V}$. The zero bias feature is understood as reflectionless tunneling \cite{marmorkos1993three} which occurs due to nonequilibrium proximity effects \cite{volkov1993proximity}, as first observed in \cite{kastalsky1991observation}. Reference \cite{marmorkos1993three} shows that the sign of the feature changes with interface transparency, with a dip indicating a high transparency case.
Given that our sample is in the ballistic regime, the most plausible origin for the finite bias subgap structures, on the other hand, is interference effects associated with the cavity.

\begin{figure}[ht]%
\includegraphics[width=\columnwidth]{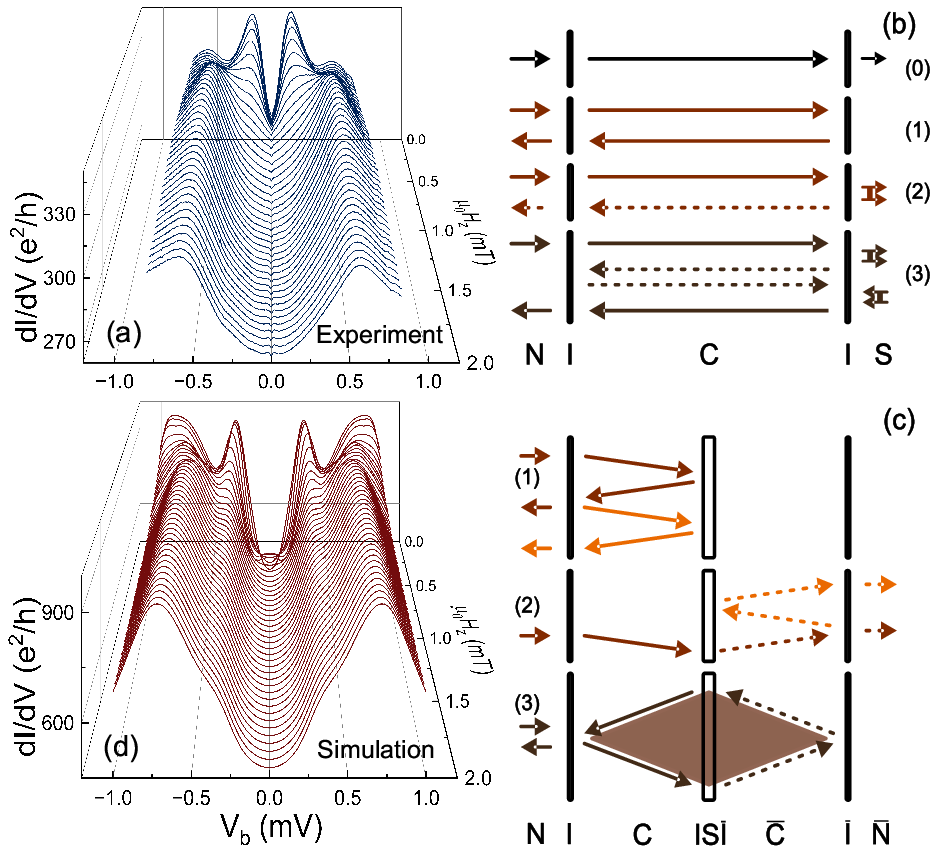}
\caption{(a) $H_{z}$ dependence of d$I$/d$V$ vs $V_{b}$  from 0 to $2\;\text{mT}$ in steps of $50\;\mu\text{T}$ (with self-gating background removed; see main text). The device length is $1000\;\text{nm}$. (b) Schematic of the 4 possible transport processes in the simplified one dimensional picture. (c) Extension of schematic processes to two dimensions. Process (1) and (2) are open trajectories with no magnetic field dependence, whereas the enclosed area of process (3) couples strongly to $H_z$ (see text.)  (d) Model curves of d$I$/d$V$ vs $V_{b}$ as a function of $H_{z}$, and corresponding to the results of a, with parameter $\gamma = 0.5$ (a phenomenological parameter describing the transparency of the interface; see text for details).}%
\label{Fig3}%
\end{figure}

In order to provide more insight and a broader basis for elaborating a model to describe these, we first investigate their experimental dependence on a small perpendicular-to-plane magnetic field ($H_{z}$). 
Fig.\ref{Fig3}(a) give the d$I$/d$V$ vs $V_b$ for various $H_{z}$ ranging from 0 to $2\;\text{mT}$ in steps of $50\;\mu\text{T}$. Fascinatingly, while the peak at $\sim600\;\mu\text{V}$ is robust to these magnetic fields, the one at $\sim180\;\mu\text{V}$ is washed out for fields above $\sim500\;\mu\text{T}$. This suggests a distinct origin for each of the two peaks.

%--------------------------------------------------------------Fig3
%-------------------------------------------------------------------

We now present a minimal phenomenological model of our device as a ballistic cavity (C) of length $L$ sandwiched between a normal metal (N) and a superconductor (S), and show that this is a sufficient description to capture the main features of the experiments.

Let us first consider an overly simplistic 1D  scenario as schematically shown in Fig.3(b). An electron entering the device from the N contact can experience 4 distinct processes: It can be transmitted into the S as an electron (process 0);  reflect as an electron from the C-S interface (process 1); Andreev reflect as a hole from the C-S interface, simultaneously injecting a Cooper pair into S, and with the hole leaving the cavity back into the normal metal (Process 2); or starting as in process 2, the hole can reflect back through the cavity, causing a second Andreev reflection, with the new electron ultimately leaving the cavity to the N contact (Process 3).     

Process (0) can be neglected at bias voltages below $\Delta$ due to absence of single particle density of states in the superconductor.

Process (1) is a back-scattering process\cite{beenakker1991quantum} returning the electron to its original reservoir, and is thus a negative contribution to the total conductance. This contribution depends on the bias voltage because the energy of the injected ballistic electron will change its interference condition.  With finite $V_{b}$ applied across the ballistic cavity, a carrier acquires a dynamical phase of $E L / \hbar v_F$\cite{nazarov2009quantum}, where $E =  e V_{b}$. The resulting resonance condition for the measured conductance as a function of bias voltage is  
\begin{equation}
    2 \frac{E L}{\hbar v_F} = (2 n+1) \pi , \quad n \in \mathbb{Z}\ ,
   \label{ee}
\end{equation}
for a double barrier system. \cite{lesovik1997nonlinearity, nazarov2009quantum}.

Process 2 is similar to process 1 with the incorporation of an Andreev reflection. It therefore has the same resonance condition (eq. 1). However, since each electron participating in this process, produces a Cooper pair in the superconductor, and returns a hole to the normal metal, the net charge transport is 2$e$. This results in a positive contribution to the enhanced conductance.

Process 3 involves two Andreev reflections, first creating and then destroying a Copper-pair, resulting in zero net charge transport. This implies the following resonance condition for this process: 
\begin{equation}
    4\frac{E L}{\hbar v_F}-2\cos^{-1}\left(\frac{E}{\Delta}\right) = 2 n \pi , \quad n \in\mathbb{Z}\ .
   \label{eh}
\end{equation}
Here, $-\cos^{-1}\left(\frac{E}{\Delta}\right)$ is the additional phase gain at the superconductor due to an Andreev reflection \cite{nazarov2009quantum}.

The above descriptions seen in Fig. 3(b) are of course only the first order of each process, which can be repeated to form higher order processes. 

The total conductance of the device (as a function of bias) can now be determined within a Landauer-Buttiker formalism using scattering matrix theory\cite{lesovik1997nonlinearity} to determine a weighted sum of the contributions of the above processes, where the relative weight of the contribution for each process is determined by cavity-superconductor interface transparency. 

Our actual device is, however, not a one dimensional system. This introduces two new aspects which must be incorporated into a realistic model: that electrons will be injected with a distribution of angles $\theta$ with respect to the interface ($\theta = 0$ is normal to the interface), and the reality that the interface transparency at the cavity-superconductor interface is not homogeneous.  

The effect of electrons being injected at angles other than $\theta = 0$ requires subtle examination. Figure 3c sketches processes 1, 2 and 3 for a finite $\theta$ (Process 0 is ignored as it can be neglected in the sub-gap regime.) Here we use a folded-space representation (See Supplemental Material at \cite{supplementalmaterial} for section S2A) where electron parts of the trajectory are drawn to the left and hole parts to the right. The darkest colored part of each process is the first order, and lighter colored arrows show the extension to second order. The representation highlights a separation of the processes into two distinct categories. 

Processes 1 and 2 yield open trajectories, i.e., do not result into an electron returning to its original position. In contrast, as a result of the retro-reflective nature of Andreev processes, process 3 creates a closed trajectory, in which the electron returns to its origin, and can thus interfere with itself.

This distinction between open and closed trajectories is key to understanding the magnetic field dependence observed in the experiment. For the small $H_{z}$ used in the experiments, any geometric correction from Lorentz force to the ballistic trajectories is completely negligible. Any processes involving open-loop trajectories are thus unaffected by magnetic field\cite{de1964magnetic, beenakker1988boundary}.

Closed-loop trajectories on the other hand, accumulate an Aharonov-Bohm phase $\phi_\text{AB}$ proportional to the flux enclosed by the trajectory, or equivalently to the shaded area in Fig. 3(c). From this area, we find $\phi_\text{AB}  = 2\frac{e}{\hbar}  L^2 \tan(\theta) \mu_0 H_z $.

Moreover, the interaction of the ballistic carriers with the Meissner screening current at the superconducting interface leads to a Doppler effect \cite{rohlfing2009doppler}. The charge conjugated particle resulting from Andreev reflection, gains an additional momentum of $\hbar k_\text{S}$ = $e \mu_0 H_z d$ \cite{tkachov2005andreev} along the interface of the superconductor, where $d$ = 4$\;\mu \text{m}$ is the width of the superconducting lead. This contributes an additional phase shift of ${\phi_\text{SC}=2 \boldsymbol{k_\text{S}}\boldsymbol{L} = 2 \frac{e}{\hbar}  Ld \tan(\theta)  \mu_0 H_z}$. The total effect of the magnetic field on the closed trajectories of process 3 is the sum: $\phi_\text{SC}+\phi_\text{AB}$ 

A further implication of the Doppler shift is that with each subsequent Andreev reflection the kinetic energy of the charge conjugated particle changes by $\delta E =- \hbar\boldsymbol{k_\text{S}}\boldsymbol{v}_F = - \sin(\theta) e v_F \mu_0 H_z d$. The cumulative energy gain after a few Andreev reflections exceeds $\Delta$ of the superconductor, and the particle escapes into the single particle density of states of the superconductor. This provides a cut-off for the maximum order of closed loop trajectories. Given the $\theta$ dependence of $\delta E$, the cut-off mechanism is stronger for larger angles.

To model the behavior of the resonances in the differential conductance
at finite magnetic field, we need to impose an inhomogeneous
superconducting interface in our model. It is assumed to be composed of
patches with lower transparency (dominated by processes (1) and (2)) and
higher transparency (dominated by process (3)). Only the combination of
the two types of patches is able to qualitatively reproduce the
experimental data. The reason is that each type of patch describes the
emergence and field dependence of one of the two resonance peaks. This
observation clearly evidences that the cavity-superconductor interface should be
inhomogeneous, and quite transparent in the experiment. We introduce a
single adjustable parameter into our model, $\gamma$, which gives the
ratio of the total area of each type of patch.

Combining the above considerations gives a full description of the system. To determine the number of injection angles to consider, we estimate the number of available ballistic transport modes ($N_{mode}$) based on a quantum point contact picture ${N_{mode} = \text{Int}[k_{F}W/\pi]}$ \cite{beenakker1991quantum}. With ${W = 5.5\;\mu \text{m}}$ and ${k_{F} = 2.2\times10^{8}\; \text{m}^{-1}}$ this yields $N_{mode}$ = 385. The total conductance ($G$)  \cite{beenakker1997random} is then given by
\begin{equation}
G= \frac{4e^2}{h} \sum_\theta (1-\gamma)\,|s_{eh}^{(\text{op})}|^2 +\gamma\,|s_{eh}^{(\text{cl})}|^2\ ,
\label{Condunctance}
\end{equation}
where we sum over a distribution of 385 angles,  $s_{eh}^{(i)}$ is the electron to hole wave amplitude of charge transport for open (i = op) and closed (i=cl) trajectories. A thorough discussion on the summation in equation 3 is given in Supplemental Material \cite{supplementalmaterial} section S2B. This provides an intuitive understanding of the emergence of the double peak structure from individual trajectories. 

With this, we calculate the differential conductance as a function of $V_{b}$. The results for $\gamma = 0.5$ are presented in \Fig{Fig3} (d) for the subgap regime. The calculated curve for zero magnetic field captures the double peak structure at finite bias. Moreover, as in the experiment, increasing $H_{z}$ suppresses the inner bias peak while the outer bias peak remains. For the numerical calculation we restrict the summation angle to a realistic upper limit of ballistic trajectories given by the ${l = 3\;\mu \text{m}}$, which corresponds to a $\theta \approx 70^\circ $. Including larger incident angles has no effect on the bias dependent subgap features, but produces numerical noise. The absolute value of the measured and modeled differential conduction differ, due to finite contact resistance at the metal-semiconductor interfaces which is ignored in the model. 
\\

%--------------------------------------------------------------Fig4
\begin{figure}[ht]%\label{fig:conductance}%
\includegraphics[width=\columnwidth]{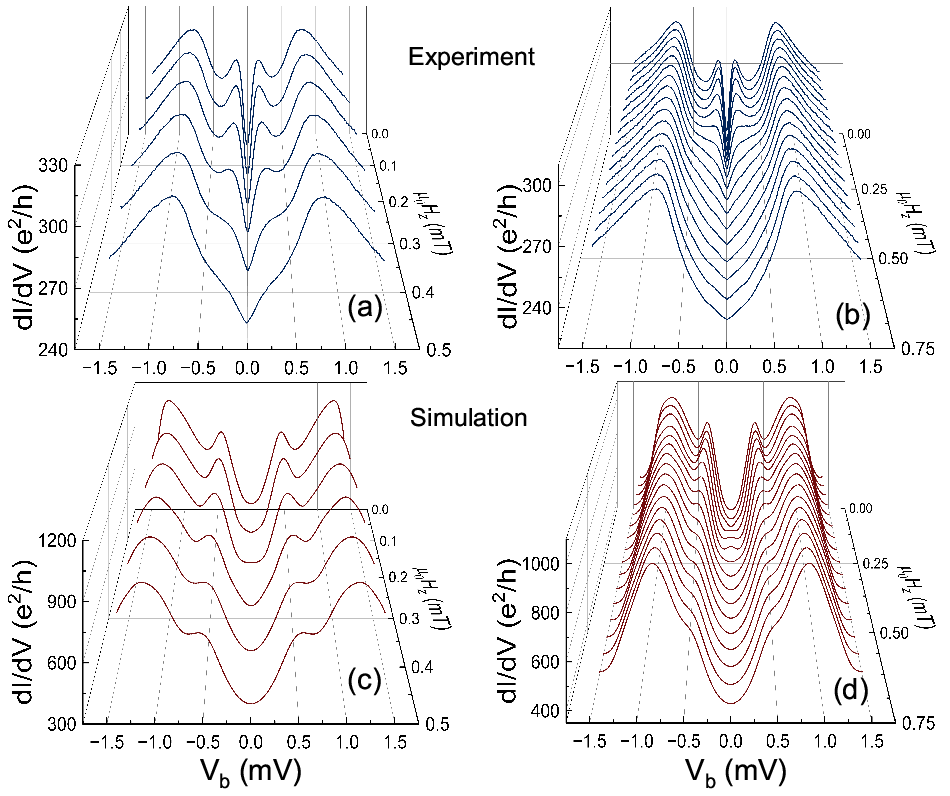}
\caption{Two additional devices. The top row shows measured data for a device with length $650\;\text{nm}$ (a) and $860\;\text{nm}$ (b). (c) and (d) show the corresponding model calculations. A value of $\gamma = 0.4$ is used in both cases.}
\label{Fig4}%
\end{figure}
%-------------------------------------------------------------------

In order to confirm the ubiquitousness of the above phenomenology, we examine two additional devices produced on a separate wafer, and with shorter lengths ($L = 650$ and $860\;\text{nm}$ respectively), with the experimental results shown in Fig. 4(a) and (b). Qualitatively, the results are similar to those in fig. 3, with a zero bias feature, and two distinct finite bias peaks, an inner with a strong decay in magnetic field, and an outer one which is robust to the field. 

Some quantitative differences can however be seen. The position of the outer peak (coming from open trajectories) moves to higher bias for shorter junctions as expected from eq.1 which shows that at the resonance condition, E and L are inversely proportional. The position of the inner peak (closed trajectories) is much less dependent on device length, as the dominant term in the resonance condition of eq. 2 is the Andreev phase.

The other quantitative differences relate to the amplitude of the features, with the inner peak having lower amplitude in both of these samples, than in that of figure 2. At the same time, in contrast with the first device, the zero bias feature in these two samples dips below the normal state conductance (see Supplemental Material at \cite{supplementalmaterial} for figure S2). As mentioned above, the zero bias feature is not included in our model, but is already understood \cite{marmorkos1993three}. The change in sign indicates that these two samples have an overall lower transparency than that in figure 2.

All of these features relating to the finite bias peaks are captured in the model curves shown in Fig. 4(c) and (d), which were produced using the same equations as figure 3(d), changing the device length, and adjusting the free parameter, $\gamma$ from 0.5 to 0.4. This corresponds to an increase in the fraction of the area with lower transparency, and is thus fully consistent with the observed behavior of the zero bias features as well as with the inner peak being of lower amplitude than the outer peak, in contrast to the first device.       

\section*{Conclusion}

Established descriptions of Andreev, and by extension Josephson, processes focus only on the nature of a single point contact at one interface between a normal material and a superconductor. That is to say that they treat the device as a zero-dimensional Sharvin type injection point to a one-dimensional transport system. While these models have been extremely successful, the  rich phenomenology described here shows that they are incomplete when faced with devices in the mesoscopic transport regime. The physics in this regime can only be captured by a model using a realistic device geometry and thus treating the cavity as a two-dimensional object whose interfaces are one-dimensional objects. Not only is such a 2D treatment essential to capture the effects of interference between ballistic trajectories, but it also reveals the role of a Doppler shift in the reflection in a more direct way than previously observed. As most relevant experiments are in the mesoscopic transport regime, a proper understanding of the physics resulting from these considerations is key to both designing, and interpreting transport experiments aimed at studying interaction between superconductors and topologically non-trivial materials.

\section*{Acknowledgments} 
\begin{acknowledgments}

The authors thank W. Beugeling, Y. Lu and F. S. Bergeret for useful discussions. The work was supported by the Free State of Bavaria (the Institute for Topological Insulators) [L.W.M.], the Deutsche Forschungsgemeinschaft (INST93/1034-1 FUGG) [C.G., L.W.M.] and (SFB 1170, 258499086) [F.D., M.P.S., E.M.H., B.T., C.G., L.W.M.], and the W\"urzburg-Dresden Cluster of Excellence on Complexity and Topology in Quantum Matter (EXC 2147, 390858490) [F.D., E.M.H., B.T., L.W.M.].
\end{acknowledgments}

\section{Author contributions}

M.P.S., T.M.K, C.G. and L.W.M. conceived and planned the experiment. L.L. and D.C. grew the material. P.M. fabricated the devices. P.M., S.M., M.P.S. and C.G. executed the measurements. M.K. and F.D. did the theoretical modelling. All authors contributed to discussion, analysis, understanding of the results and participated in the preparation of the manuscript.
\color{black}
\section{Competing interests} The authors declare that they have no competing interests.

\section{Data availability}
All data needed to evaluate the conclusions in the paper are present in the paper and/or the Supplementary Materials. Additional data related to this paper may be requested from the authors.

\renewcommand{\thefigure}{S\arabic{figure}}
\setcounter{figure}{0}

\section*{Supplementary Material}

\subsection*{S1: Experiment}

\textbf{A: Carrier density from Shubnikov - de Haas measurement:}  
We use Shubnikov - de Haas (SdH) measurement to estimate the carrier density of the device for which the results are presented in the main text Fig. 2 and 3(a). Figure S1 shows conductance ($G$) in units of $e^{2}/h$ vs inverse of applied perpendicular-to-plane magnetic field (H$_{z}$). Magnetic field values of the indicated red points are read from the data with their corresponding filling factor to extract SdH period from the linear fit (in blue). This gives a charier density of 7.8$\times$10$^{11}$ cm$^{-2}$. \\

%--------------------------------------------------------------SupplFig1
\begin{figure}[h]%
\includegraphics[width=\columnwidth]{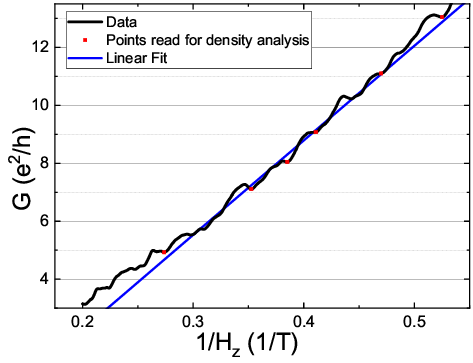}
\caption{ Conductance in units of $e^{2}/h$ vs inverse of applied magnetic field H$_{z}$ measured at 4 K. Data points in red are read by eye as minimums, and used to extract SdH period using a linear fit shown in blue.
}%
\label{SupplFig1}%
\end{figure}
%-------------------------------------------------------------------

\textbf{B: Normalized differential conductance:}

%--------------------------------------------------------------SupplFig2
\begin{figure}[h]%
\includegraphics[width=\columnwidth]{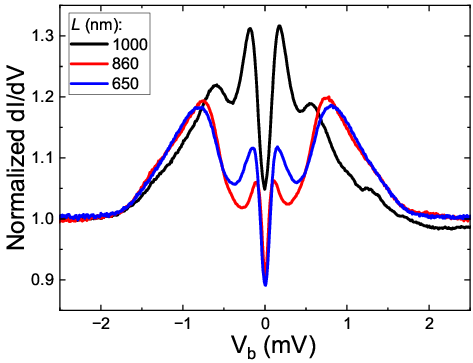}
\caption{  Normalized d$I$/d$V$ measurement at base temperature w.r.t zero-bias conductance above the critical temperature of MoRe, after removing the background tilt from the raw data, for devices with $L$ = 1000 nm (in black), 860 nm (in red) and 650 nm (in blue).
}%
\label{SupplFig2}%
\end{figure}
%-----------------------------------------------------------------

To compare the three devices with varied length $L$, we look at their d$I$/d$V$ curves at base temperature, in each case normalized to the normal state zero-bias conductance measured above the critical temperature of MoRe, and after removing the background tilt from the raw data, as in the main text. Figure S2 shows the quantitative variation between device with $L$ = 1000 nm and other two devices with $L$ = 860 and 650 nm, which are discussed in the main text.  \\

\subsection*{S2: Model}

\textbf{A: Folded-space representation}

%--------------------------------------------------------------SupplFig3
\begin{figure}[h]%
\includegraphics[width=\columnwidth]{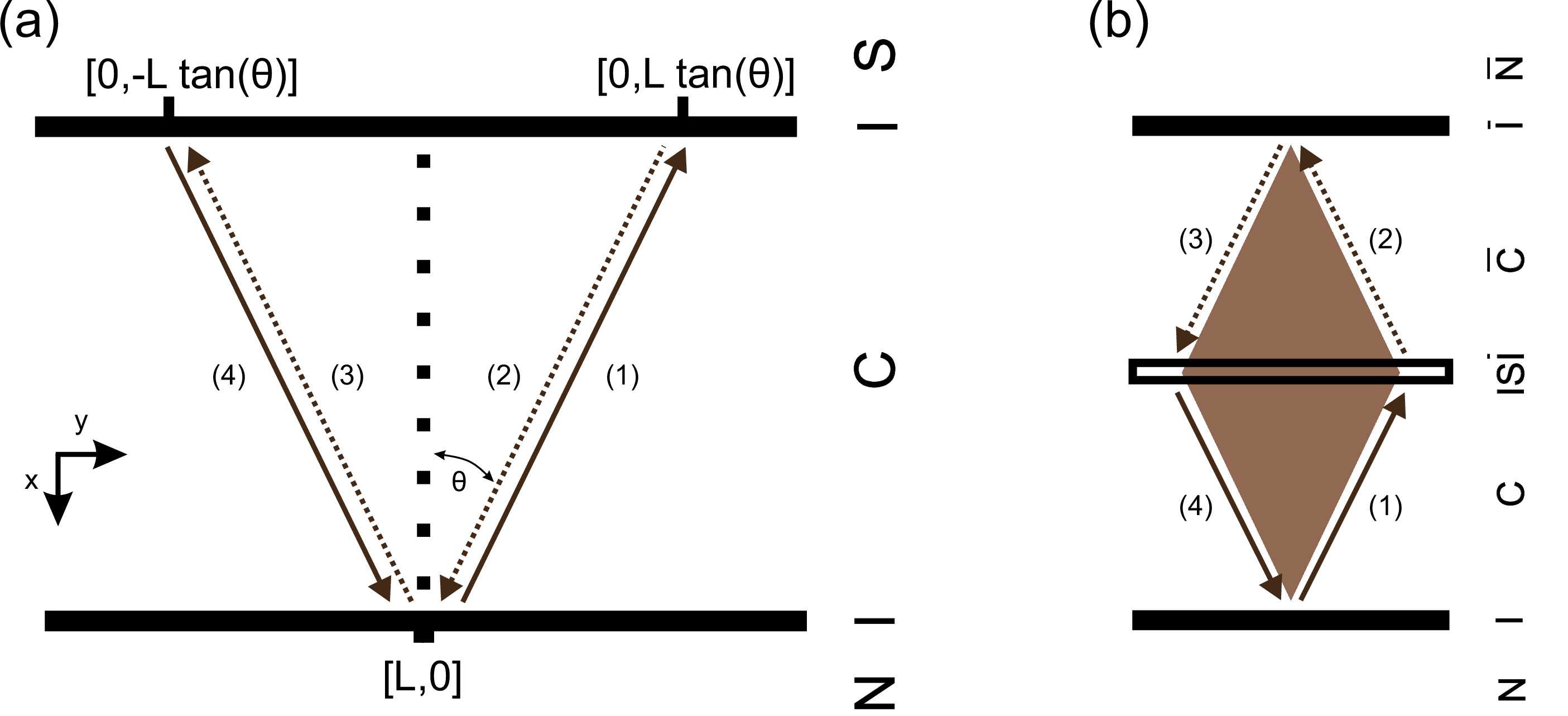}
%\includesvg[width=1.0\columnwidth]{ParityDiagram}
\caption{Schematic diagram of Process (3) discussed in main text in a real space picture (a) and parity folded picture (b). Parity space is mirrored in respect to the superconducting interface.  
}%
\label{SupplFig3}%
\end{figure}
%-----------------------------------------------------------------

To show processes that from closed loop trajectories, it is beneficial to use a different diagrammatic representation than a real space depiction of trajectories.
The real space depiction of Process (3), as discussed in the main text, is shown in Fig. \ref{SupplFig3} (a).
As the goal of these figure is to illustrate the Andreev reflection, it is useful to view the system as two different, fictitious, regions, one for the electron and one for the hole, to avoid drawing trajectories on top of each other. This is done in fig. S3b, where the lower half shows the electron part of the trajectory, whereas the upper half shows, as a mirror reflection, the hole part of the trajectory.   
\\
This representation is particularly useful in the context of the Aharonov-Bohm effect, because the magnetic phase gain of a traveling particle is invariant to a simultaneous chance of both parity ($x \to -x$ in respect to the superconducting interface) and charge. The magnetic phase gain calculated over each path (1-4) results into the same value, where
\begin{equation}
    \phi_{(1)} = \frac{e}{h} \mu_0 H_z L^2 \int_0^1 r \text{d}r = \frac{(-e)}{h} \mu_0 H_z L^2 \int_1^0 r \text{d}r    =  \phi_{(2)} 
\end{equation}
Here, $r$ is trajectory parametrization from $[x_0,y_0] \to [x_1,y_1]$ with $r \in [0,1]$. 
 Understanding that the Andreev processes by nature is retro reflective and charge converting, we can draw a picture of a single particle tracing the closed loop of the folded-space representation in Fig. \ref{SupplFig3} (b). Parity space here is labeled with a bar on top of the letter. The trajectory of this fictitious particle encloses a shaded area, and from Stokes' theorem gains a magnetic phase proportional to magnetic field times enclosed area.
Using the presented equivalent representation provides an intuitive way of understanding Aharonov-Bohm in the context of Andreev processes.  \\
%\newpage

\textbf{B: Angle dependent differential conductance}  

%--------------------------------------------------------------SupplFig4
\begin{figure}[h]%
\includegraphics[width=\columnwidth]{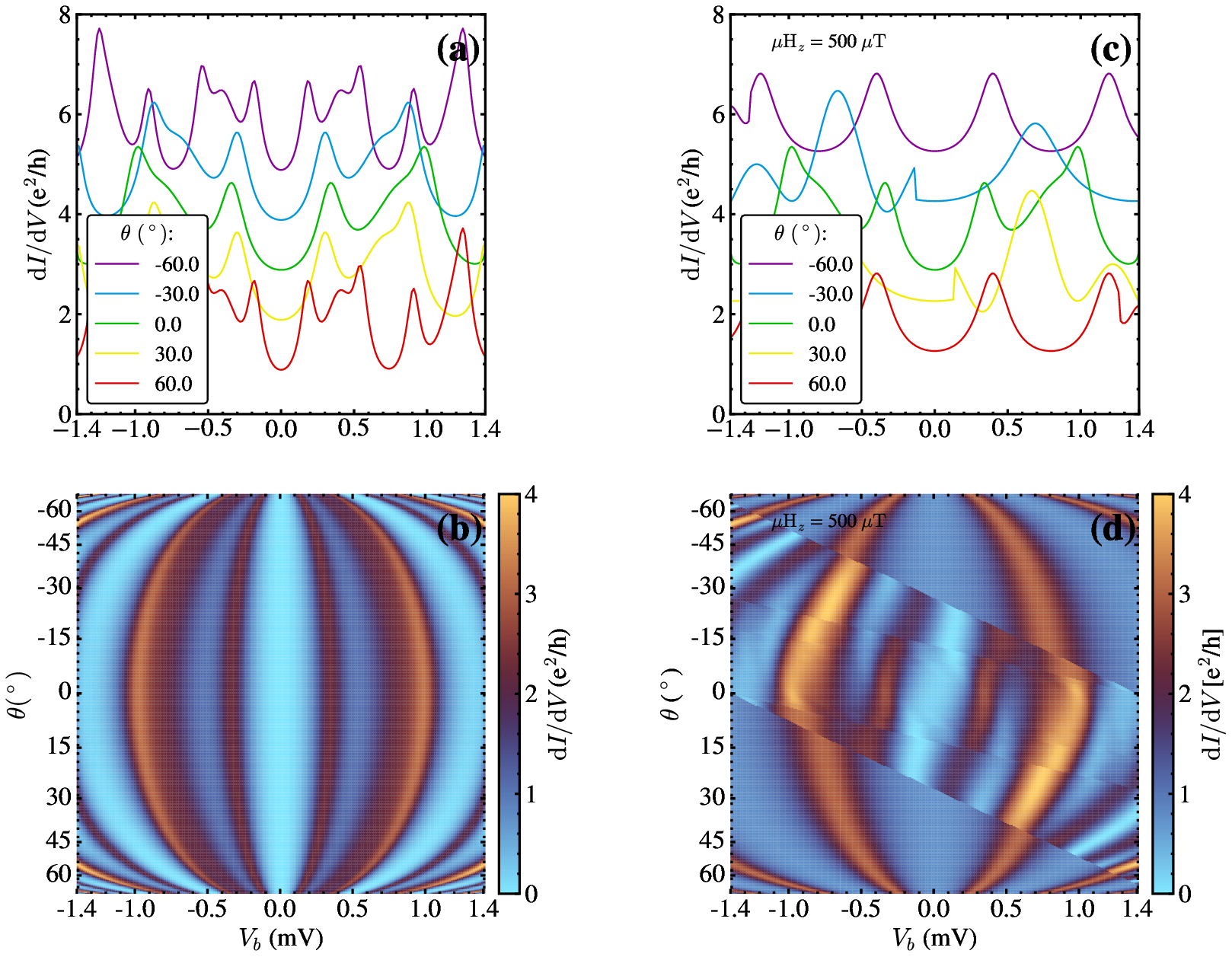}
\caption{ (a,c) Waterfall plot of d$I$/d$V$ vs $V_b$ for modes injected into the cavity at various angles (see legend) for 0 (a) and ${500\ \mu \text{T}}$ (c) respectively. For clarity, with respect to the $60^\circ$ the other curves are incrementally offset by $1 e^2/h$ vertically. (b,d) Color-map of d$I$/d$V$ in relation to angle of injection $\theta$ and applied $V_b$ for 0 (b) and ${500\ \mu \text{T}}$ (d) respectively.  Distinct feature become visible due to the interference pattern originating from resonance equation (1) and (2), given in the main text. In (d) visible straight lines of discontinuity are created by truncation due to Doppler-shift. In addition the inner resonant features of closed loop trajectories start to tilt due to the additional magnetic phase. Note that the y-axis on (b,d) use cosine scaling.      
}%
\label{SupplFig4}%
\end{figure}
%-----------------------------------------------------------------

 This section expands on the incorporation of angular summation as per equation (3), which generates the simulated differential conduction in the main text. Given that the experiment involves many conducting channels with a multitude of interfering trajectories, an averaging over all resonances might naively be expected to produce a featureless differential conductance in the subgap regime.
 However, the experimental results show a clear double peak structure. To examine the summation over trajectories, we first begin with looking at differential conductance of individual modes injected at different angles.

A waterfall plot of d$I$/d$V$ vs applied $V_b$ for individual angles is presented in Fig. \ref{SupplFig4} (a) for modes that are injected into the cavity at $\theta = 0,\pm 30^\circ,\pm 60^\circ$ with an incremental vertical offset of $1 e^2/h$ starting with $60^\circ$ curve. Although the bias dependent position of individual peaks and valleys changes with angles, the behavior is actually systematic. This is evident when looking at higher angular resolution data in the color-map representation of differential conductance in relation to applied bias and the angle of injection, as seen in Fig. \ref{SupplFig4} (b). The scaling of the y-axis uses a standard cosine dependence in line with the distribution of injected angles. The result of averaging over all possible trajectories at a given bias voltage is simply the summation of a vertical slice of Fig. \ref{SupplFig4} (b). The pronounced vertically bowed golden bands that fill the majority of the picture neatly explain the distinct emergence of the double peak structure also visible in the experiment.

This approach is also applicable in a magnetic field and explains the suppression of the inner peak. For an applied magnetic field of ${500\ \mu \text{T}}$ the corresponding differential conductance plot as in 
\ref{SupplFig4} (a) is presented in Fig. \ref{SupplFig4} (c) and similar color-map is given in Fig. \ref{SupplFig4} (d). The diagonal bands forming discontinuities visible in figure \ref{SupplFig4} (d) result from the effect of Doppler-shift. As discussed in the main text the contribution of trajectories are cut off if the number of collisions with the superconductor causes the energy gain to exceed the gap.
The integer number of collision required to meet this criteria is angle dependent. The slope of these discontinuities is then directly proportional to $H_z$. Going closer to the origin of the figure, more discontinuities due to higher order truncation appear.

In addition the inner resonant bands of Fig. \ref{SupplFig4}(b) coming from closed loop trajectories start to tilt.
Here one can see, by adding the magnetic phase $\phi_{\text{SC}}+\phi_{\text{AB}}\propto \tan(\theta)$ trajectories that were before at resonance shift to larger or smaller applied bias, to compensate the additional phase they gain per roundtrip inside the cavity. Note that due to the tangential dependence of magnetic phase and the presence of positive and negative angles Fig. \ref{SupplFig4}(d) is symmetric with $180^\circ$ rotation, while the mirror symmetry ($\theta\to-\theta, V_b \to -V_b$) is broken. This is in accordance with a break of time reversal symmetry by magnetic field.

When taking both Doppler-effect and magnetic phase into consideration, the resonant bands corresponding to the closed trajectories of equation (2) are either  truncated or twisted resulting into a suppression of the inner peak.


\section{References}


\begin{thebibliography}{30}%
\makeatletter
\providecommand \@ifxundefined [1]{%
 \@ifx{#1\undefined}
}%
\providecommand \@ifnum [1]{%
 \ifnum #1\expandafter \@firstoftwo
 \else \expandafter \@secondoftwo
 \fi
}%
\providecommand \@ifx [1]{%
 \ifx #1\expandafter \@firstoftwo
 \else \expandafter \@secondoftwo
 \fi
}%
\providecommand \natexlab [1]{#1}%
\providecommand \enquote  [1]{``#1''}%
\providecommand \bibnamefont  [1]{#1}%
\providecommand \bibfnamefont [1]{#1}%
\providecommand \citenamefont [1]{#1}%
\providecommand \href@noop [0]{\@secondoftwo}%
\providecommand \href [0]{\begingroup \@sanitize@url \@href}%
\providecommand \@href[1]{\@@startlink{#1}\@@href}%
\providecommand \@@href[1]{\endgroup#1\@@endlink}%
\providecommand \@sanitize@url [0]{\catcode `\\12\catcode `\$12\catcode
  `\&12\catcode `\#12\catcode `\^12\catcode `\_12\catcode `\%12\relax}%
\providecommand \@@startlink[1]{}%
\providecommand \@@endlink[0]{}%
\providecommand \url  [0]{\begingroup\@sanitize@url \@url }%
\providecommand \@url [1]{\endgroup\@href {#1}{\urlprefix }}%
\providecommand \urlprefix  [0]{URL }%
\providecommand \Eprint [0]{\href }%
\providecommand \doibase [0]{https://doi.org/}%
\providecommand \selectlanguage [0]{\@gobble}%
\providecommand \bibinfo  [0]{\@secondoftwo}%
\providecommand \bibfield  [0]{\@secondoftwo}%
\providecommand \translation [1]{[#1]}%
\providecommand \BibitemOpen [0]{}%
\providecommand \bibitemStop [0]{}%
\providecommand \bibitemNoStop [0]{.\EOS\space}%
\providecommand \EOS [0]{\spacefactor3000\relax}%
\providecommand \BibitemShut  [1]{\csname bibitem#1\endcsname}%
\let\auto@bib@innerbib\@empty
%</preamble>
\bibitem [{\citenamefont {Andreev}(1964)}]{Andreev1964}%
  \BibitemOpen
  \bibfield  {author} {\bibinfo {author} {\bibfnamefont {A.~F.}\ \bibnamefont
  {Andreev}},\ }\bibfield  {title} {\bibinfo {title} {The thermal conductivity
  of the intermediate state in superconductors},\ }\href@noop {} {\bibfield
  {journal} {\bibinfo  {journal} {Soviet Physics JETP}\ }\textbf {\bibinfo {volume}
  {19}},\ \bibinfo {pages} {1228} (\bibinfo {year} {1964})},\ \bibinfo {note}
  {[\textit{Zh. Eksp. Teor. Fiz.} \textbf{46}, 1823 (1964)]}\BibitemShut
  {NoStop}%
\bibitem [{\citenamefont {Taylor}\ and\ \citenamefont
  {Burstein}(1963)}]{taylor1963excess}%
  \BibitemOpen
  \bibfield  {author} {\bibinfo {author} {\bibfnamefont {B.}~\bibnamefont
  {Taylor}}\ and\ \bibinfo {author} {\bibfnamefont {E.}~\bibnamefont
  {Burstein}},\ }\bibfield  {title} {\bibinfo {title} {Excess currents in
  electron tunneling between superconductors},\ }\href@noop {} {\bibfield
  {journal} {\bibinfo  {journal} {Physical Review Letters}\ }\textbf {\bibinfo
  {volume} {10}},\ \bibinfo {pages} {14} (\bibinfo {year} {1963})}\BibitemShut
  {NoStop}%
\bibitem [{\citenamefont {Notarys}\ and\ \citenamefont
  {Mercereau}(1973)}]{notarys1973proximity}%
  \BibitemOpen
  \bibfield  {author} {\bibinfo {author} {\bibfnamefont {H.}~\bibnamefont
  {Notarys}}\ and\ \bibinfo {author} {\bibfnamefont {J.}~\bibnamefont
  {Mercereau}},\ }\bibfield  {title} {\bibinfo {title} {Proximity effect bridge
  and superconducting microcircuitry},\ }\href@noop {} {\bibfield  {journal}
  {\bibinfo  {journal} {Journal of Applied Physics}\ }\textbf {\bibinfo
  {volume} {44}},\ \bibinfo {pages} {1821} (\bibinfo {year}
  {1973})}\BibitemShut {NoStop}%
\bibitem [{\citenamefont {Akimenko}\ \emph {et~al.}(1976)\citenamefont
  {Akimenko}, \citenamefont {Solov’ev},\ and\ \citenamefont
  {Yanson}}]{akimenko1976ac}%
  \BibitemOpen
  \bibfield  {author} {\bibinfo {author} {\bibfnamefont {A.}~\bibnamefont
  {Akimenko}}, \bibinfo {author} {\bibfnamefont {V.}~\bibnamefont
  {Solov’ev}},\ and\ \bibinfo {author} {\bibfnamefont {I.}~\bibnamefont
  {Yanson}},\ }\bibfield  {title} {\bibinfo {title} {ac josephson current as a
  function of voltage},\ }\href@noop {} {\bibfield  {journal} {\bibinfo
  {journal} {Soviet Journal of Low Temperature Physics}\ }\textbf {\bibinfo
  {volume} {2}},\ \bibinfo {pages} {238} (\bibinfo {year} {1976})},\ \bibinfo
  {note} {[\textit{Fiz. Nizk. Temp.} \textbf{2}, 480 (1976)]}\BibitemShut
  {NoStop}%
\bibitem [{\citenamefont {Klapwijk}\ \emph {et~al.}(1977)\citenamefont
  {Klapwijk}, \citenamefont {Sepers},\ and\ \citenamefont
  {Mooij}}]{klapwijk1977regimes}%
  \BibitemOpen
  \bibfield  {author} {\bibinfo {author} {\bibfnamefont {T.}~\bibnamefont
  {Klapwijk}}, \bibinfo {author} {\bibfnamefont {M.}~\bibnamefont {Sepers}},\
  and\ \bibinfo {author} {\bibfnamefont {J.}~\bibnamefont {Mooij}},\ }\bibfield
   {title} {\bibinfo {title} {Regimes in the behavior of superconducting
  microbridges},\ }\href@noop {} {\bibfield  {journal} {\bibinfo  {journal}
  {Journal of Low Temperature Physics}\ }\textbf {\bibinfo {volume} {27}},\
  \bibinfo {pages} {801} (\bibinfo {year} {1977})}\BibitemShut {NoStop}%
\bibitem [{\citenamefont {Octavio}\ \emph {et~al.}(1978)\citenamefont
  {Octavio}, \citenamefont {Skocpol},\ and\ \citenamefont
  {Tinkham}}]{octavio1978nonequilibrium}%
  \BibitemOpen
  \bibfield  {author} {\bibinfo {author} {\bibfnamefont {M.}~\bibnamefont
  {Octavio}}, \bibinfo {author} {\bibfnamefont {W.}~\bibnamefont {Skocpol}},\
  and\ \bibinfo {author} {\bibfnamefont {M.}~\bibnamefont {Tinkham}},\
  }\bibfield  {title} {\bibinfo {title} {Nonequilibrium-enhanced supercurrents
  in short superconducting weak links},\ }\href@noop {} {\bibfield  {journal}
  {\bibinfo  {journal} {Physical Review B}\ }\textbf {\bibinfo {volume} {17}},\
  \bibinfo {pages} {159} (\bibinfo {year} {1978})}\BibitemShut {NoStop}%
\bibitem [{\citenamefont {Gubankov}\ \emph {et~al.}(1981)\citenamefont
  {Gubankov}, \citenamefont {Koshelets},\ and\ \citenamefont
  {Ovsyannikov}}]{gubankov1981excess}%
  \BibitemOpen
  \bibfield  {author} {\bibinfo {author} {\bibfnamefont {V.}~\bibnamefont
  {Gubankov}}, \bibinfo {author} {\bibfnamefont {V.}~\bibnamefont
  {Koshelets}},\ and\ \bibinfo {author} {\bibfnamefont {G.}~\bibnamefont
  {Ovsyannikov}},\ }\bibfield  {title} {\bibinfo {title} {Excess current in
  variable-thickness superconducting microbridges},\ }\href@noop {} {\bibfield
  {journal} {\bibinfo  {journal} {Soviet Journal of Low Temperature Physics}\
  }\textbf {\bibinfo {volume} {7}},\ \bibinfo {pages} {135} (\bibinfo {year}
  {1981})},\ \bibinfo {note} {[\textit{Fiz. Nizk. Temp.} \textbf{7}, 277
  (1981)]}\BibitemShut {NoStop}%
\bibitem [{\citenamefont {Blonder}\ \emph {et~al.}(1982)\citenamefont
  {Blonder}, \citenamefont {Tinkham},\ and\ \citenamefont
  {Klapwijk}}]{blonder1982transition}%
  \BibitemOpen
  \bibfield  {author} {\bibinfo {author} {\bibfnamefont {G.}~\bibnamefont
  {Blonder}}, \bibinfo {author} {\bibfnamefont {M.}~\bibnamefont {Tinkham}},\
  and\ \bibinfo {author} {\bibfnamefont {T.}~\bibnamefont {Klapwijk}},\
  }\bibfield  {title} {\bibinfo {title} {Transition from metallic to tunneling
  regimes in superconducting microconstrictions: Excess current, charge
  imbalance, and supercurrent conversion},\ }\href@noop {} {\bibfield
  {journal} {\bibinfo  {journal} {Physical Review B}\ }\textbf {\bibinfo
  {volume} {25}},\ \bibinfo {pages} {4515} (\bibinfo {year}
  {1982})}\BibitemShut {NoStop}%
\bibitem [{\citenamefont {Fu}\ and\ \citenamefont
  {Kane}(2008)}]{fu2008superconducting}%
  \BibitemOpen
  \bibfield  {author} {\bibinfo {author} {\bibfnamefont {L.}~\bibnamefont
  {Fu}}\ and\ \bibinfo {author} {\bibfnamefont {C.~L.}\ \bibnamefont {Kane}},\
  }\bibfield  {title} {\bibinfo {title} {Superconducting proximity effect and
  Majorana fermions at the surface of a topological insulator},\ }\href@noop {}
  {\bibfield  {journal} {\bibinfo  {journal} {Physical Review Letters}\
  }\textbf {\bibinfo {volume} {100}},\ \bibinfo {pages} {096407} (\bibinfo
  {year} {2008})}\BibitemShut {NoStop}%
\bibitem [{\citenamefont {Qi}\ \emph {et~al.}(2010)\citenamefont {Qi},
  \citenamefont {Hughes},\ and\ \citenamefont {Zhang}}]{qi2010chiral}%
  \BibitemOpen
  \bibfield  {author} {\bibinfo {author} {\bibfnamefont {X.-L.}\ \bibnamefont
  {Qi}}, \bibinfo {author} {\bibfnamefont {T.~L.}\ \bibnamefont {Hughes}},\
  and\ \bibinfo {author} {\bibfnamefont {S.-C.}\ \bibnamefont {Zhang}},\
  }\bibfield  {title} {\bibinfo {title} {Chiral topological superconductor from
  the quantum Hall state},\ }\href@noop {} {\bibfield  {journal} {\bibinfo
  {journal} {Physical Review B—Condensed Matter and Materials Physics}\
  }\textbf {\bibinfo {volume} {82}},\ \bibinfo {pages} {184516} (\bibinfo
  {year} {2010})}\BibitemShut {NoStop}%
\bibitem [{\citenamefont {Van~Ostaay}\ \emph {et~al.}(2011)\citenamefont
  {Van~Ostaay}, \citenamefont {Akhmerov},\ and\ \citenamefont
  {Beenakker}}]{van2011spin}%
  \BibitemOpen
  \bibfield  {author} {\bibinfo {author} {\bibfnamefont {J.}~\bibnamefont
  {Van~Ostaay}}, \bibinfo {author} {\bibfnamefont {A.}~\bibnamefont
  {Akhmerov}},\ and\ \bibinfo {author} {\bibfnamefont {C.}~\bibnamefont
  {Beenakker}},\ }\bibfield  {title} {\bibinfo {title} {Spin-triplet
  supercurrent carried by quantum Hall edge states through a Josephson
  junction},\ }\href@noop {} {\bibfield  {journal} {\bibinfo  {journal}
  {Physical Review B—Condensed Matter and Materials Physics}\ }\textbf
  {\bibinfo {volume} {83}},\ \bibinfo {pages} {195441} (\bibinfo {year}
  {2011})}\BibitemShut {NoStop}%
\bibitem [{\citenamefont {Rowell}\ and\ \citenamefont
  {Feldmann}(1968)}]{rowell1968excess}%
  \BibitemOpen
  \bibfield  {author} {\bibinfo {author} {\bibfnamefont {J.}~\bibnamefont
  {Rowell}}\ and\ \bibinfo {author} {\bibfnamefont {W.}~\bibnamefont
  {Feldmann}},\ }\bibfield  {title} {\bibinfo {title} {Excess currents in
  superconducting tunnel junctions},\ }\href@noop {} {\bibfield  {journal}
  {\bibinfo  {journal} {Physical Review}\ }\textbf {\bibinfo {volume} {172}},\
  \bibinfo {pages} {393} (\bibinfo {year} {1968})}\BibitemShut {NoStop}%
\bibitem [{\citenamefont {Octavio}\ \emph {et~al.}(1983)\citenamefont
  {Octavio}, \citenamefont {Tinkham}, \citenamefont {Blonder},\ and\
  \citenamefont {Klapwijk}}]{octavio1983subharmonic}%
  \BibitemOpen
  \bibfield  {author} {\bibinfo {author} {\bibfnamefont {M.}~\bibnamefont
  {Octavio}}, \bibinfo {author} {\bibfnamefont {M.}~\bibnamefont {Tinkham}},
  \bibinfo {author} {\bibfnamefont {G.}~\bibnamefont {Blonder}},\ and\ \bibinfo
  {author} {\bibfnamefont {T.}~\bibnamefont {Klapwijk}},\ }\bibfield  {title}
  {\bibinfo {title} {Subharmonic energy-gap structure in superconducting
  constrictions},\ }\href@noop {} {\bibfield  {journal} {\bibinfo  {journal}
  {Physical Review B}\ }\textbf {\bibinfo {volume} {27}},\ \bibinfo {pages}
  {6739} (\bibinfo {year} {1983})}\BibitemShut {NoStop}%
\bibitem [{\citenamefont {Pandey}\ \emph {et~al.}(2019)\citenamefont {Pandey},
  \citenamefont {Kraft}, \citenamefont {Krupke}, \citenamefont {Beckmann},\
  and\ \citenamefont {Danneau}}]{pandey2019andreev}%
  \BibitemOpen
  \bibfield  {author} {\bibinfo {author} {\bibfnamefont {P.}~\bibnamefont
  {Pandey}}, \bibinfo {author} {\bibfnamefont {R.}~\bibnamefont {Kraft}},
  \bibinfo {author} {\bibfnamefont {R.}~\bibnamefont {Krupke}}, \bibinfo
  {author} {\bibfnamefont {D.}~\bibnamefont {Beckmann}},\ and\ \bibinfo
  {author} {\bibfnamefont {R.}~\bibnamefont {Danneau}},\ }\bibfield  {title}
  {\bibinfo {title} {Andreev reflection in ballistic normal
  metal/graphene/superconductor junctions},\ }\href@noop {} {\bibfield
  {journal} {\bibinfo  {journal} {Physical Review B}\ }\textbf {\bibinfo
  {volume} {100}},\ \bibinfo {pages} {165416} (\bibinfo {year}
  {2019})}\BibitemShut {NoStop}%
\bibitem [{\citenamefont {Nazarov}\ and\ \citenamefont
  {Blanter}(2009)}]{nazarov2009quantum}%
  \BibitemOpen
  \bibfield  {author} {\bibinfo {author} {\bibfnamefont {Y.~V.}\ \bibnamefont
  {Nazarov}}\ and\ \bibinfo {author} {\bibfnamefont {Y.~M.}\ \bibnamefont
  {Blanter}},\ }\href@noop {} {\emph {\bibinfo {title} {Quantum transport:
  Introduction to Nanoscience}}}\ (\bibinfo  {publisher} {Cambridge university
  press},\ \bibinfo {year} {2009})\BibitemShut {NoStop}%
\bibitem [{\citenamefont {Bendias}\ \emph {et~al.}(2018)\citenamefont
  {Bendias}, \citenamefont {Shamim}, \citenamefont {Herrmann}, \citenamefont
  {Budewitz}, \citenamefont {Shekhar}, \citenamefont {Leubner}, \citenamefont
  {Kleinlein}, \citenamefont {Bocquillon}, \citenamefont {Buhmann},\ and\
  \citenamefont {Molenkamp}}]{bendias2018high}%
  \BibitemOpen
  \bibfield  {author} {\bibinfo {author} {\bibfnamefont {K.}~\bibnamefont
  {Bendias}}, \bibinfo {author} {\bibfnamefont {S.}~\bibnamefont {Shamim}},
  \bibinfo {author} {\bibfnamefont {O.}~\bibnamefont {Herrmann}}, \bibinfo
  {author} {\bibfnamefont {A.}~\bibnamefont {Budewitz}}, \bibinfo {author}
  {\bibfnamefont {P.}~\bibnamefont {Shekhar}}, \bibinfo {author} {\bibfnamefont
  {P.}~\bibnamefont {Leubner}}, \bibinfo {author} {\bibfnamefont
  {J.}~\bibnamefont {Kleinlein}}, \bibinfo {author} {\bibfnamefont
  {E.}~\bibnamefont {Bocquillon}}, \bibinfo {author} {\bibfnamefont
  {H.}~\bibnamefont {Buhmann}},\ and\ \bibinfo {author} {\bibfnamefont {L.~W.}\
  \bibnamefont {Molenkamp}},\ }\bibfield  {title} {\bibinfo {title} {High
  mobility HgTe microstructures for quantum spin Hall studies},\ }\href@noop {}
  {\bibfield  {journal} {\bibinfo  {journal} {Nano Letters}\ }\textbf {\bibinfo
  {volume} {18}},\ \bibinfo {pages} {4831} (\bibinfo {year}
  {2018})}\BibitemShut {NoStop}%
\bibitem [{\citenamefont {Mandal}\ \emph {et~al.}(2022)\citenamefont {Mandal},
  \citenamefont {Taufertsh{\"o}fer}, \citenamefont {Lunczer}, \citenamefont
  {Stehno}, \citenamefont {Gould},\ and\ \citenamefont
  {Molenkamp}}]{mandal2022finite}%
  \BibitemOpen
  \bibfield  {author} {\bibinfo {author} {\bibfnamefont {P.}~\bibnamefont
  {Mandal}}, \bibinfo {author} {\bibfnamefont {N.}~\bibnamefont
  {Taufertsh{\"o}fer}}, \bibinfo {author} {\bibfnamefont {L.}~\bibnamefont
  {Lunczer}}, \bibinfo {author} {\bibfnamefont {M.~P.}\ \bibnamefont {Stehno}},
  \bibinfo {author} {\bibfnamefont {C.}~\bibnamefont {Gould}},\ and\ \bibinfo
  {author} {\bibfnamefont {L.~W.}\ \bibnamefont {Molenkamp}},\ }\bibfield
  {title} {\bibinfo {title} {Finite field transport response of a dilute
  magnetic topological insulator-based Josephson junction},\ }\href@noop {}
  {\bibfield  {journal} {\bibinfo  {journal} {Nano Letters}\ }\textbf {\bibinfo
  {volume} {22}},\ \bibinfo {pages} {3557} (\bibinfo {year}
  {2022})}\BibitemShut {NoStop}%
\bibitem [{sup()}]{supplementalmaterial}%
  \BibitemOpen
  \bibfield  {title} {\bibinfo {title} {See Supplemental Material [URL] for additional information on carrier density, comparison of normalized differential conductance for three devices with various length, folded-space representation, and details of angular summation in the model}\
  }\href@noop {} {\ }\BibitemShut {NoStop}%
\bibitem [{\citenamefont {Beugeling}\ \emph {et~al.}(2025)\citenamefont
  {Beugeling}, \citenamefont {Bayer}, \citenamefont {Berger}, \citenamefont
  {B{\"o}ttcher}, \citenamefont {Bovkun}, \citenamefont {Fuchs}, \citenamefont
  {Hofer}, \citenamefont {Shamim}, \citenamefont {Siebert}, \citenamefont
  {Wang} \emph {et~al.}}]{beugeling2025kdotpy}%
  \BibitemOpen
  \bibfield  {author} {\bibinfo {author} {\bibfnamefont {W.}~\bibnamefont
  {Beugeling}}, \bibinfo {author} {\bibfnamefont {F.}~\bibnamefont {Bayer}},
  \bibinfo {author} {\bibfnamefont {C.}~\bibnamefont {Berger}}, \bibinfo
  {author} {\bibfnamefont {J.}~\bibnamefont {B{\"o}ttcher}}, \bibinfo {author}
  {\bibfnamefont {L.}~\bibnamefont {Bovkun}}, \bibinfo {author} {\bibfnamefont
  {C.}~\bibnamefont {Fuchs}}, \bibinfo {author} {\bibfnamefont
  {M.}~\bibnamefont {Hofer}}, \bibinfo {author} {\bibfnamefont
  {S.}~\bibnamefont {Shamim}}, \bibinfo {author} {\bibfnamefont
  {M.}~\bibnamefont {Siebert}}, \bibinfo {author} {\bibfnamefont {L.-X.}\
  \bibnamefont {Wang}}, \emph {et~al.},\ }\bibfield  {title} {\bibinfo {title}
  {kdotpy: k{\textperiodcentered} p theory on a lattice for simulating
  semiconductor band structures},\ }\href@noop {} {\bibfield  {journal}
  {\bibinfo  {journal} {SciPost Physics Codebases}\ ,\ \bibinfo {pages} {047}}
  (\bibinfo {year} {2025})}\BibitemShut {NoStop}%
\bibitem [{\citenamefont {Kristensen}\ \emph {et~al.}(2000)\citenamefont
  {Kristensen}, \citenamefont {Bruus}, \citenamefont {Hansen}, \citenamefont
  {Jensen}, \citenamefont {Lindelof}, \citenamefont {Marckmann}, \citenamefont
  {Nyg{\aa}rd}, \citenamefont {S{\o}rensen}, \citenamefont {Beuscher},
  \citenamefont {Forchel} \emph {et~al.}}]{kristensen2000bias}%
  \BibitemOpen
  \bibfield  {author} {\bibinfo {author} {\bibfnamefont {A.}~\bibnamefont
  {Kristensen}}, \bibinfo {author} {\bibfnamefont {H.}~\bibnamefont {Bruus}},
  \bibinfo {author} {\bibfnamefont {A.}~\bibnamefont {Hansen}}, \bibinfo
  {author} {\bibfnamefont {J.}~\bibnamefont {Jensen}}, \bibinfo {author}
  {\bibfnamefont {P.}~\bibnamefont {Lindelof}}, \bibinfo {author}
  {\bibfnamefont {C.}~\bibnamefont {Marckmann}}, \bibinfo {author}
  {\bibfnamefont {J.}~\bibnamefont {Nyg{\aa}rd}}, \bibinfo {author}
  {\bibfnamefont {C.}~\bibnamefont {S{\o}rensen}}, \bibinfo {author}
  {\bibfnamefont {F.}~\bibnamefont {Beuscher}}, \bibinfo {author}
  {\bibfnamefont {A.}~\bibnamefont {Forchel}}, \emph {et~al.},\ }\bibfield
  {title} {\bibinfo {title} {Bias and temperature dependence of the 0.7
  conductance anomaly in quantum point contacts},\ }\href@noop {} {\bibfield
  {journal} {\bibinfo  {journal} {Physical Review B}\ }\textbf {\bibinfo
  {volume} {62}},\ \bibinfo {pages} {10950} (\bibinfo {year}
  {2000})}\BibitemShut {NoStop}%
\bibitem [{\citenamefont {Marmorkos}\ \emph {et~al.}(1993)\citenamefont
  {Marmorkos}, \citenamefont {Beenakker},\ and\ \citenamefont
  {Jalabert}}]{marmorkos1993three}%
  \BibitemOpen
  \bibfield  {author} {\bibinfo {author} {\bibfnamefont {I.}~\bibnamefont
  {Marmorkos}}, \bibinfo {author} {\bibfnamefont {C.}~\bibnamefont
  {Beenakker}},\ and\ \bibinfo {author} {\bibfnamefont {R.}~\bibnamefont
  {Jalabert}},\ }\bibfield  {title} {\bibinfo {title} {Three signatures of
  phase-coherent Andreev reflection},\ }\href@noop {} {\bibfield  {journal}
  {\bibinfo  {journal} {Physical Review B}\ }\textbf {\bibinfo {volume} {48}},\
  \bibinfo {pages} {2811} (\bibinfo {year} {1993})}\BibitemShut {NoStop}%
\bibitem [{\citenamefont {Volkov}\ \emph {et~al.}(1993)\citenamefont {Volkov},
  \citenamefont {Zaitsev},\ and\ \citenamefont
  {Klapwijk}}]{volkov1993proximity}%
  \BibitemOpen
  \bibfield  {author} {\bibinfo {author} {\bibfnamefont {A.}~\bibnamefont
  {Volkov}}, \bibinfo {author} {\bibfnamefont {A.}~\bibnamefont {Zaitsev}},\
  and\ \bibinfo {author} {\bibfnamefont {T.}~\bibnamefont {Klapwijk}},\
  }\bibfield  {title} {\bibinfo {title} {Proximity effect under nonequilibrium
  conditions in double-barrier superconducting junctions},\ }\href@noop {}
  {\bibfield  {journal} {\bibinfo  {journal} {Physica C: Superconductivity}\
  }\textbf {\bibinfo {volume} {210}},\ \bibinfo {pages} {21} (\bibinfo {year}
  {1993})}\BibitemShut {NoStop}%
\bibitem [{\citenamefont {Kastalsky}\ \emph {et~al.}(1991)\citenamefont
  {Kastalsky}, \citenamefont {Kleinsasser}, \citenamefont {Greene},
  \citenamefont {Bhat}, \citenamefont {Milliken},\ and\ \citenamefont
  {Harbison}}]{kastalsky1991observation}%
  \BibitemOpen
  \bibfield  {author} {\bibinfo {author} {\bibfnamefont {A.}~\bibnamefont
  {Kastalsky}}, \bibinfo {author} {\bibfnamefont {A.}~\bibnamefont
  {Kleinsasser}}, \bibinfo {author} {\bibfnamefont {L.}~\bibnamefont {Greene}},
  \bibinfo {author} {\bibfnamefont {R.}~\bibnamefont {Bhat}}, \bibinfo {author}
  {\bibfnamefont {F.}~\bibnamefont {Milliken}},\ and\ \bibinfo {author}
  {\bibfnamefont {J.}~\bibnamefont {Harbison}},\ }\bibfield  {title} {\bibinfo
  {title} {Observation of pair currents in superconductor-semiconductor
  contacts},\ }\href@noop {} {\bibfield  {journal} {\bibinfo  {journal}
  {Physical Review Letters}\ }\textbf {\bibinfo {volume} {67}},\ \bibinfo
  {pages} {3026} (\bibinfo {year} {1991})}\BibitemShut {NoStop}%
\bibitem [{\citenamefont {Beenakker}\ and\ \citenamefont {van
  Houten}(1991)}]{beenakker1991quantum}%
  \BibitemOpen
  \bibfield  {author} {\bibinfo {author} {\bibfnamefont {C.}~\bibnamefont
  {Beenakker}}\ and\ \bibinfo {author} {\bibfnamefont {H.}~\bibnamefont {van
  Houten}},\ }\bibfield  {title} {\bibinfo {title} {Quantum transport in
  semiconductor nanostructures},\ }in\ \href@noop {} {\emph {\bibinfo
  {booktitle} {Solid state physics}}},\ Vol.~\bibinfo {volume} {44}\ (\bibinfo
  {publisher} {Elsevier},\ \bibinfo {year} {1991})\ pp.\ \bibinfo {pages}
  {1--228}\BibitemShut {NoStop}%
\bibitem [{\citenamefont {Lesovik}\ \emph {et~al.}(1997)\citenamefont
  {Lesovik}, \citenamefont {Fauchere},\ and\ \citenamefont
  {Blatter}}]{lesovik1997nonlinearity}%
  \BibitemOpen
  \bibfield  {author} {\bibinfo {author} {\bibfnamefont {G.~B.}\ \bibnamefont
  {Lesovik}}, \bibinfo {author} {\bibfnamefont {A.~L.}\ \bibnamefont
  {Fauchere}},\ and\ \bibinfo {author} {\bibfnamefont {G.}~\bibnamefont
  {Blatter}},\ }\bibfield  {title} {\bibinfo {title} {Nonlinearity in
  normal-metal--superconductor transport: Scattering-matrix approach},\
  }\href@noop {} {\bibfield  {journal} {\bibinfo  {journal} {Physical Review
  B}\ }\textbf {\bibinfo {volume} {55}},\ \bibinfo {pages} {3146} (\bibinfo
  {year} {1997})}\BibitemShut {NoStop}%
\bibitem [{\citenamefont {De~Gennes}\ and\ \citenamefont
  {Tinkham}(1964)}]{de1964magnetic}%
  \BibitemOpen
  \bibfield  {author} {\bibinfo {author} {\bibfnamefont {P.}~\bibnamefont
  {De~Gennes}}\ and\ \bibinfo {author} {\bibfnamefont {M.}~\bibnamefont
  {Tinkham}},\ }\bibfield  {title} {\bibinfo {title} {Magnetic behavior of very
  small superconducting particles},\ }\href@noop {} {\bibfield  {journal}
  {\bibinfo  {journal} {Physics Physique Fizika}\ }\textbf {\bibinfo {volume}
  {1}},\ \bibinfo {pages} {107} (\bibinfo {year} {1964})}\BibitemShut {NoStop}%
\bibitem [{\citenamefont {Beenakker}\ and\ \citenamefont
  {Van~Houten}(1988)}]{beenakker1988boundary}%
  \BibitemOpen
  \bibfield  {author} {\bibinfo {author} {\bibfnamefont {C.}~\bibnamefont
  {Beenakker}}\ and\ \bibinfo {author} {\bibfnamefont {H.}~\bibnamefont
  {Van~Houten}},\ }\bibfield  {title} {\bibinfo {title} {Boundary scattering
  and weak localization of electrons in a magnetic field},\ }\href@noop {}
  {\bibfield  {journal} {\bibinfo  {journal} {Physical Review B}\ }\textbf
  {\bibinfo {volume} {38}},\ \bibinfo {pages} {3232} (\bibinfo {year}
  {1988})}\BibitemShut {NoStop}%
\bibitem [{\citenamefont {Rohlfing}\ \emph {et~al.}(2009)\citenamefont
  {Rohlfing}, \citenamefont {Tkachov}, \citenamefont {Otto}, \citenamefont
  {Richter}, \citenamefont {Weiss}, \citenamefont {Borghs},\ and\ \citenamefont
  {Strunk}}]{rohlfing2009doppler}%
  \BibitemOpen
  \bibfield  {author} {\bibinfo {author} {\bibfnamefont {F.}~\bibnamefont
  {Rohlfing}}, \bibinfo {author} {\bibfnamefont {G.}~\bibnamefont {Tkachov}},
  \bibinfo {author} {\bibfnamefont {F.}~\bibnamefont {Otto}}, \bibinfo {author}
  {\bibfnamefont {K.}~\bibnamefont {Richter}}, \bibinfo {author} {\bibfnamefont
  {D.}~\bibnamefont {Weiss}}, \bibinfo {author} {\bibfnamefont
  {G.}~\bibnamefont {Borghs}},\ and\ \bibinfo {author} {\bibfnamefont
  {C.}~\bibnamefont {Strunk}},\ }\bibfield  {title} {\bibinfo {title} {Doppler
  shift in Andreev reflection from a moving superconducting condensate in
  Nb/InAs Josephson junctions},\ }\href@noop {} {\bibfield  {journal} {\bibinfo
   {journal} {Physical Review B—Condensed Matter and Materials Physics}\
  }\textbf {\bibinfo {volume} {80}},\ \bibinfo {pages} {220507} (\bibinfo
  {year} {2009})}\BibitemShut {NoStop}%
\bibitem [{\citenamefont {Tkachov}\ and\ \citenamefont
  {Richter}(2005)}]{tkachov2005andreev}%
  \BibitemOpen
  \bibfield  {author} {\bibinfo {author} {\bibfnamefont {G.}~\bibnamefont
  {Tkachov}}\ and\ \bibinfo {author} {\bibfnamefont {K.}~\bibnamefont
  {Richter}},\ }\bibfield  {title} {\bibinfo {title} {Andreev magnetotransport
  in low-dimensional proximity structures: Spin-dependent conductance
  enhancement},\ }\href@noop {} {\bibfield  {journal} {\bibinfo  {journal}
  {Physical Review B—Condensed Matter and Materials Physics}\ }\textbf
  {\bibinfo {volume} {71}},\ \bibinfo {pages} {094517} (\bibinfo {year}
  {2005})}\BibitemShut {NoStop}%
\bibitem [{\citenamefont {Beenakker}(1997)}]{beenakker1997random}%
  \BibitemOpen
  \bibfield  {author} {\bibinfo {author} {\bibfnamefont {C.~W.}\ \bibnamefont
  {Beenakker}},\ }\bibfield  {title} {\bibinfo {title} {Random-matrix theory of
  quantum transport},\ }\href@noop {} {\bibfield  {journal} {\bibinfo
  {journal} {Reviews of modern physics}\ }\textbf {\bibinfo {volume} {69}},\
  \bibinfo {pages} {731} (\bibinfo {year} {1997})}\BibitemShut {NoStop}%
\end{thebibliography}
\end{document}